# Miniaturized superconducting metamaterials for radio frequencies

Cihan Kurter,[1,a)] John Abrahams,[1,2] and Steven M. Anlage[1]
[1]*Department of Physics, Center for Nanophysics and Advanced Materials, University of Maryland, College Park, Maryland 20742-4111, USA*
[2]*Department of Electrical and Computer Engineering, University of Maryland, College Park, Maryland 20742-3285, USA*

We have developed a low-loss, ultrasmall radio frequency (rf) metamaterial operating at ∼76 MHz. This miniaturized medium is made up of planar spiral elements with diameter as small as ∼λ/658 (λ is the free space wavelength), fashioned from Nb thin films on quartz substrates. The transmission data are examined below and above the superconducting transition temperature of Nb for both a single spiral and a one dimensional array. The validity of the design is tested through numerical simulations and good agreement is found. We discuss how superconductors enable such a compact design in the rf with high loaded-quality factor (in excess of 5000), which is in fact difficult to realize with ordinary metals. © *2010 American Institute of Physics*.
[doi:10.1063/1.3456524]

The wavelength of visible light is orders of magnitude larger than the size of the atoms in most dielectric media. Therefore, light is not sensitive to the details of the atomic-scale electromagnetic fields but to coarse-grained properties of the structure. Since metamaterials mimic natural materials, artificial electromagnetic structures must have elements whose dimensions are much smaller than the free space wavelength (λ) at which the metamaterial operates. Up to now, research has focused on metamaterials functioning at gigahertz or higher frequencies, since the electrical size of the inclusion of such a medium can be kept reasonably small.[1–5] Many metamaterials are based on split ring resonators (SRRs) and their variations to create subwavelength magnetically active features. The SRR geometry is typically composed of two concentric rings, each with a capacitive split, and can be understood in terms of the inductor-capacitor (*LC*) analogy. However this design calls for comparatively large dimensions for radio frequency (rf) metamaterials, since the wavelength is on the order of meters.

Although there are few studies available on rf metamaterials, they have great potential in applications such as magnetic resonance imaging (MRI) devices for noninvasive and high resolution medical imaging.[6,7] They may also improve rf antenna efficiency and directivity as well as reduce their size.[8] Microwave delay lines,[9] magnetoinductive lenses for near field imaging,[10] rf filters, and compact resonators[11] may also be enabled by rf metamaterials.

The first practical demonstration of those metamaterials was implemented by Wiltshire *et al.*[6,12,13] using a Swiss roll geometry as suggested earlier by Pendry.[14] An array of the rolls made of Cu/Kapton layers wounded around a dielectric mandrel was used in an MRI machine for guiding rf flux caused by magnetic resonance to the receiver coil in the system.[6,15] Though the design had some disadvantages, such as being lossy, three-dimensional, and nonuniform, it was well-suited for such applications.[16] A planar version of the Swiss roll geometry fabricated with a thick copper film on a dielectric substrate and excited by a microstrip transmission line was found to resonate at frequencies as low as 125 MHz. However significant and maximum losses were observed at the resonance of the effective permeability.[17] Magnetically coupled elements were also realized to not only modify the propagation of transverse electromagnetic waves but also to generate magnetoinductive waves.[18,19] This sort of radiation was examined both in planar[9,20] and bulk geometries[21,22] to gain insight about its dispersion and propagation characteristics.

Our design has two key advantages over previous work making the element size especially compact: a truly planar spiral geometry and superconducting thin films. The spiral design utilizes the *LC* analogy just like SRRs. If a single spiral element is unwound, the length of the entire winding would be ∼0.6 m which is comparable to the wavelength. Keeping the strip width small (10 μm) enables many turns in a spiral with a small outer diameter, which increases the total inductance as well as capacitance thus decreasing the fundamental resonance frequency, $f_0$. Moreover, due to the superconductivity, there is a significant contribution of kinetic inductance ($L_k$) to the total inductance.[23] $L_k$ is a measure of kinetic energy of charge carriers in a current carrying material and is intimately related to the superfluid density dependence on temperature in superconductors. $L_k$ could also be modified through application of large rf magnetic fields or dc magnetic field,[24] as long as the sample remains in the Meissner state. This kind of inductance is negligibly small in normal metals at any temperature and frequency in the rf band.

The geometrical inductance ($L_g$) of a spiral with $N$ turns, outer diameter of $D_o$ and inner diameter of $D_i$ can be approximated as:[25] $L_g = (\mu_0 N^2 D_{avg})/2[\ln(2.46/a) + 0.2a^2]$ where $D_{avg} = (D_o - D_i)/2$ and $a = (D_o - D_i)/(D_o + D_i)$. The distributed capacitance of planar spirals is given in Ref. 26 as $C(pF) = 0.035 D_o(mm) + 0.06$. As seen here, both $L_g$ and $C$ increase with $D_o$ (thus reducing the resonant frequency), however there is also a need to minimize the size of each element. For temperatures $T \ll T_c$ ($T_c$ is the superconducting transition temperature), the resonance frequency can be estimated by ignoring $L_k$ as $f = 1/(2\pi\sqrt{L_g C})$. However at tem-

a)Electronic mail: ckurter@umd.edu.

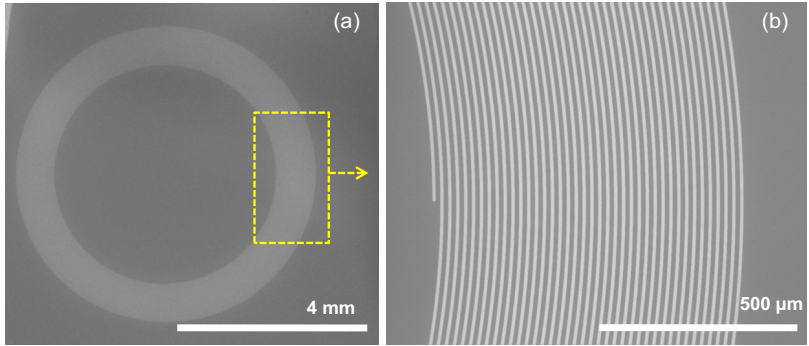

FIG. 1. (Color online) (a) SEM image of a single Nb spiral with $D_o$=6 mm, $N$=40, $w$=$s$=10 $\mu$m on a quartz substrate and (b) the SEM image of right part of the spiral marked by yellow in (a) showing the windings.

peratures close to $T_c$, $L_k$ should be added to $L_g$ to get a precise value for the resonance frequency.

Fabrication of our rf metamaterials begins with deposition of a 200 nm Nb thin film on a 350 $\mu$m thick quartz wafer, 3″ in diameter, using the rf sputtering technique. Superconductivity of the Nb thin films is tested by resistivity measurements which give a $T_c$ of ~9.24 K. Photolithography is applied to pattern the film surface into the shape of spirals having 40 turns of $w$=10 $\mu$m wide wires with $s$=10 $\mu$m spacing. The chemically reactive plasma generated by a mixture of $CF_4/O_2$ (10% $O_2$) etches the film down to the quartz to give a final shape to the spiral resonators. The quartz is diced into chips having single spirals or arrays. In Fig. 1(a), a scanning electron microscopy (SEM) image of a representative single spiral with $D_o$=6 mm is shown, while Fig. 1(b) shows the details of the same spiral.

For measurement of the rf properties, the quartz substrate with Nb spiral is sandwiched between two magnetic loops connected to an Agilent E5062A rf network analyzer. The magnetic loops ~6 mm in diameter are made by bending the stripped inner conductors and soldering them to the outer parts of the coax cables. The rf excitation is provided to the spiral resonator by the top loop. The transmission is picked up by the bottom loop and sent to the rf network analyzer. The alternating current flowing along the top loop creates a time varying magnetic field which couples to the spiral via a mutual inductance $M$ and induces an electromotive force and current. The magnetic field generated through the spiral due to the circulating current couples to the bottom loop in the same way. The inset of Fig. 2 is a simulation showing the magnetic flux lines in the sample-probe configuration done by high frequency structure simulator (HFSS). The experiments have been conducted in an evacuated probe inside a cryogenic dewar and the temperature has been precisely adjusted by a Lakeshore 340 Temperature Controller.

The transmission data $|S_{21}|$ on an element of our rf metamaterial, a single spiral with $D_o$=6 mm, $N$=40 turns and $w$=$s$=10 $\mu$m, are shown in the main panel of Fig. 2 while Nb is both in the superconducting (below $T_c$) and normal state (above $T_c$). Below $T_c$, $f_0$ appears at ~76 MHz followed by harmonics, whereas the normal state does not show any resonant features. The inset of Fig. 3 shows an HFSS simulation of current density ($J$) distribution in the spiral for $n$=1 (fundamental mode) and $n$=2 (second harmonic). The red areas indicate where the current is strong, whereas green corresponds to small values of $J$. As seen from the simulations, for $n$=1 there is a robust current distribution flowing through the middle windings, which gets weaker around the edges of the spiral. However for $n$=2, there are two counter-propagating current distributions and a large degree of cancellation. The main panel of Fig. 3 shows the calculated mutual inductance $M$ between the resonating spiral and a 6 mm diameter single-turn filamentary loop centered on the spiral, based on the Neumann integral,[27] as a function of mode number $n$. As seen in the figure, $M$ is large only for the odd modes, semiquantitatively explaining the alternating large and small amplitudes of $|S_{21}|$ resonances in Fig. 2.

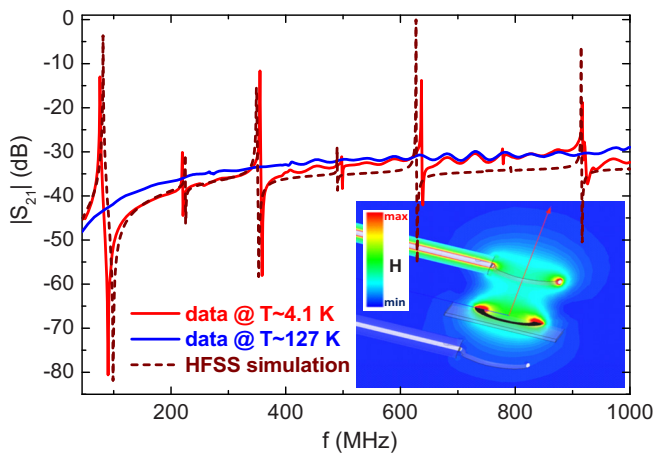

FIG. 2. (Color online) Transmission $|S_{21}|$ vs frequency for a single Nb spiral with $D_o$=6 mm measured between two rf magnetic loops at temperatures below (the solid bright red curve) and well above (the featureless solid blue curve) the $T_c$ of Nb. The dashed dark red curve is a numerical calculation done by HFSS. The inset shows the sample-loop configuration during transmission measurements as well as the distribution of the magnetic flux lines simulated by HFSS.

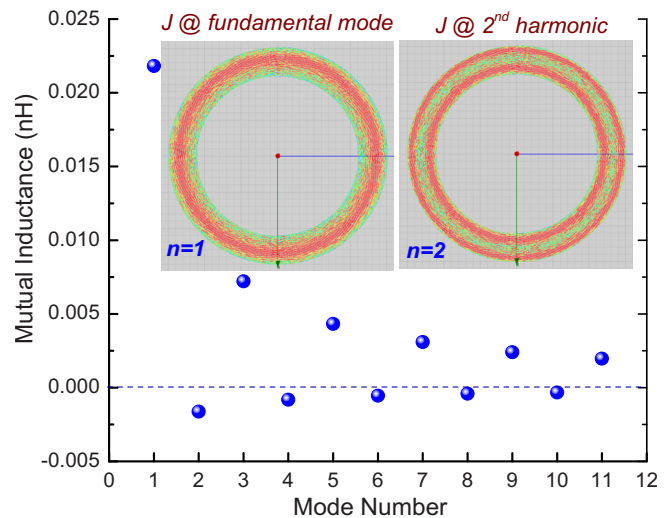

FIG. 3. (Color online) Calculation of mutual inductance $M$ between a single spiral with $D_o$=6 mm, $N$=40, $w$=$s$=10 $\mu$m and a magnetic loop with diameter 6 mm as a function of standing wave mode number in the spiral. The planes of the loop and the spiral are parallel and 3.25 mm apart. The inset shows HFSS simulations of distributed $J$ in the spiral at the fundamental mode and second harmonic.

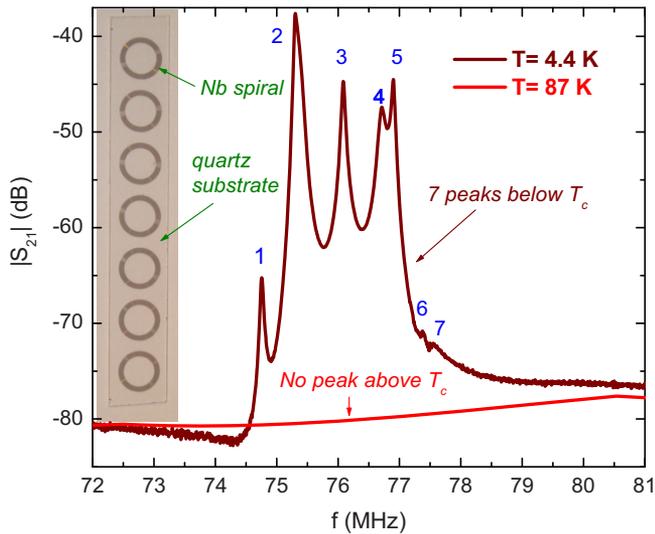

FIG. 4. (Color online) Transmission $|S_{21}|$ vs frequency for a one dimensional array of seven spirals with $D_o=6$ mm measured via magnetic loops below and above $T_c$. The inset is an optical image of the sample showing seven spirals on a quartz substrate.

Using HFSS, the spiral sample and the coax cables (magnetic loops) are modeled in detail and the simulation gives very consistent results with the experiments below $T_c$ (dashed dark red curve in Fig. 2). Note that the simulations assume the spirals are made of perfect metal, showing that the finite quality factor ($Q$) of the fundamental mode is limited by coupling and radiation loss, rather than Ohmic losses. The measured loaded $Q$ of our rf metamaterials is found to be as high as 5044 at $T=4.2$ K, depending on coupling (this value is obtained for a 27 mm separation between the two loops). Other reported $Q$ values are 60 at $f_0=21.5$ MHz for Swiss rolls,[13] 115 for capacitively loaded rings at $f_0=63.28$ MHz (Ref. 7) and roughly 143 for planar normal metal spirals at $f_0=122.5$ MHz (see Fig. 2 of Ref. 17).

Fig. 4 shows the transmission data on an array of seven equally spaced spirals with $D_o=6$ mm below and above the $T_c$ of Nb. The distance between the centers of the spirals is 7.5 mm. In this measurement the top loop is aligned with the first spiral whereas the bottom loop is aligned with the last one so that the induced magnetic field on the first resonant element by the top loop will inductively couple from one spiral to the next, forming a magnetoinductive transmission line. Although seven individual resonant features are observed in a frequency span centered on $\sim 76$ MHz below $T_c$, the resonance peaks vary in intensity based on their coupling to each other and the excitation loop. The dominant magnetic coupling is assumed to be due to nearest neighbors in the array,[22] however there is a magnetic interaction between far neighbors as well,[12] which may modify the shapes of the resonance curves.

We have demonstrated electrically small rf metamaterials with magnetically active elements as small as $\sim\lambda/658$, yet retaining a high loaded $Q\sim 5000$. These superconducting metamaterials have many advantages over their normal metal counterparts, such as minimizing the Ohmic losses, compact structure, and sensitive tuning of resonant features via temperature and magnetic field. Moreover, the design promises further reduction in resonant frequency down to 10 MHz with a modified geometry (with thinner windings) and enhanced $L_k$ as well as Josephson and vortex inductances.

This project has been supported by the U.S. Office of Naval Research through Grant No. N000140811058 and the Center for Nanophysics and Advanced Materials at the University of Maryland. We would like to thank Brian Straughn, John Hummel, and Jonathan Hood for their assistance.

[1] D. R. Smith, W. J. Padilla, D. C. Vier, S. C. Nemat-Nasser, and S. Schultz, Phys. Rev. Lett. **84**, 4184 (2000).
[2] R. A. Shelby, D. R. Smith, and S. Schultz, Science **292**, 77 (2001).
[3] S. Linden, C. Enkrich, M. Wegener, J. F. Zhou, T. Koschny, and C. M. Soukoulis, Science **306**, 1351 (2004).
[4] N. Katsarakis, T. Koschny, M. Kafesaki, E. N. Economou, and C. M. Soukoulis, Appl. Phys. Lett. **84**, 2943 (2004).
[5] M. Ricci, N. Orloff, and S. M. Anlage, Appl. Phys. Lett. **87**, 034102 (2005).
[6] M. C. K. Wiltshire, J. B. Pendry, I. R. Young, D. J. Larkman, D. J. Gilderdale, and J. V. Hajnal, Science **291**, 849 (2001).
[7] M. J. Freire, R. Marques, and L. Jelinek, Appl. Phys. Lett. **93**, 231108 (2008).
[8] R. W. Ziolkowski and A. Erentok, IEEE Trans. Antennas Propag. **54**, 2113 (2006).
[9] M. J. Freire, R. Marques, F. Medina, M. A. G. Laso, and F. Martin, Appl. Phys. Lett. **85**, 4439 (2004).
[10] M. J. Freire and R. Marques, Appl. Phys. Lett. **86**, 182505 (2005).
[11] N. Engheta, IEEE Antennas Wireless Propag. Lett. **1**, 10 (2002).
[12] M. C. K. Wiltshire, E. Shamonina, I. R. Young, and L. Solymar, J. Appl. Phys. **95**, 4488 (2004).
[13] M. C. K. Wiltshire, J. B. Pendry, W. Williams, and J. V. Hajnal, J. Phys. Condens. Matter **19**, 456216 (2007).
[14] J. B. Pendry, A. J. Holden, D. J. Robbins, and W. J. Stewart, IEEE Trans. Microwave Theory Tech. **47**, 2075 (1999).
[15] M. C. K. Wiltshire, Phys. Status Solidi B **244**, 1227 (2007).
[16] M. C. K. Wiltshire, J. V. Hajnal, J. B. Pendry, D. J. Edwards, and C. J. Stevens, Opt. Express **11**, 709 (2003).
[17] S. Massaoudi and I. Huynen, Microwave Opt. Technol. Lett. **50**, 1945 (2008).
[18] E. Shamonina, V. A. Kalinin, K. H. Ringhofer, and L. Solymar, J. Appl. Phys. **92**, 6252 (2002).
[19] E. Shamonina, V. A. Kalinin, K. H. Ringhofer, and L. Solymar, Electron. Lett. **38**, 371 (2002).
[20] R. R. A. Syms, I. R. Young, and L. Solymar, J. Phys. D: Appl. Phys. **39**, 3945 (2006).
[21] E. Shamonina and L. Solymar, J. Phys. D: Appl. Phys. **37**, 362 (2004).
[22] M. C. K. Wiltshire, E. Shamonina, I. R. Young, and L. Solymar, Electron. Lett. **39**, 215 (2003).
[23] M. C. Ricci and S. M. Anlage, Appl. Phys. Lett. **88**, 264102 (2006).
[24] M. C. Ricci, H. Xu, R. Prozorov, A. P. Zhuravel, A. V. Ustinov, and S. M. Anlage, IEEE Trans. Appl. Supercond. **17**, 918 (2007).
[25] S. S. Mohan, M. M. Hershenson, S. P. Boyd, and T. H. Lee, IEEE J. Solid-State Circuits **34**, 1419 (1999).
[26] Z. Jiang, P. S. Excell, and Z. M. Hejazi, IEEE Trans. Microwave Theory Tech. **45**, 139 (1997).
[27] J. D. Jackson, *Classical Electrodynamics*, 3rd ed. (Wiley, New York, 1998).